\documentclass[preprint,5p,twocolumn]{m-elsarticle}
\usepackage{amsmath}
\usepackage{amssymb}
\usepackage{epsfig}
\usepackage{color}
\definecolor{bl}{cmyk}{1,1,0.3,0}
\definecolor{gr}{rgb}{0,0.35,0.1}
\definecolor{new}{rgb}{0.21,0.43,0.5}

\usepackage{hyperref}
\hypersetup{colorlinks=true,urlcolor=new,linkcolor=new,citecolor=new}
\newcommand{\dif}{\textrm{d}}
\newcommand{\Mpl}{\textrm{M}_{\textrm{pl}}^2}
\newcommand{\Lam}{\Lambda^{4}}
\newcommand{\refp}[1]{(\ref{#1})}
\begin{document}

%%%%%%%%%%%%%%%%%%%%%%%%%%%%%%%%%%%%%%%%%%%%%%%%%%
\begin{frontmatter}
%%%%%%%%%%%%%%%%%%%%%%%%%%%%%%%%%%%%%%%%%%%%%%%%%%
\title{\textsf{Black hole solutions in massive
    gravity
%\\
%{\small A contribution to dark matter or anti-gravity}
}}
%%%
\author[ulb]{Michael V. Bebronne}
\author[ulb,INR]{Peter G. Tinyakov}
\address[ulb]{Service de Physique Th\'eorique, Universit\'e Libre de Bruxelles (ULB),\\ CP225, boulevard du Triomphe, B-1050 Bruxelles, Belgium.}
\address[INR]{Institute for Nuclear Research of the Russian Academy of Sciences,\\ 60th October Anniversary Prospect, 7a, 117312 Moscow, Russia.}
%%%
\begin{abstract}
The static vacuum spherically symmetric solutions in massive gravity
are obtained both analytically and numerically. The solutions depend
on two parameters (integration constants): the mass $M$ (or,
equivalently, the Schwarzschild radius), and an additional parameter,
the ``scalar charge'' $S$. At zero value of $S$ and positive mass the
standard Schwarzschild black hole solutions are recovered.  Depending
on the parameters of the model and the signs of $M$ and $S$, the
solutions may or may not have horizon. Those with the horizon describe
modified black holes provided they are stable against small
perturbations.  In the analytically solvable example, the modified
black hole solutions may have both attractive and repulsive
(anti-gravitating) behavior at large distances.  At intermediate
distances the gravitational potential of a modified black hole may
mimics the presence of dark matter. Modified black hole solutions are
also found numerically in more realistic massive gravity models which
are attractors of the cosmological evolution. 

\end{abstract}

%%%%%%%%%%%%%%%%%%%%%%%%%%%%%%%%%%%%%%%%%%%%%%%%%%
\end{frontmatter}
%%%%%%%%%%%%%%%%%%%%%%%%%%%%%%%%%%%%%%%%%%%%%%%%%%

\section{Introduction} 
\label{sc:intro}

Recent advances in observational cosmology
\cite{Perlmutter:1998np,Dunkley:2008ie,AdelmanMcCarthy:2007wh} have
revived interest in alternative theories of gravity in which the
gravitational interaction is modified in the infrared domain and which
could --- potentially --- explain the accelerated expansion of the
Universe without introducing the dark energy and matter
components. Theoretical consistency and existing experimental data
impose severe constraints on such models. Different approaches to the
problem have been discussed in the literature (see, e.g.,
Refs.~\cite{Milgrom:1983pn,Bekenstein:2004ne,Dvali:2000hr,Gregory:2000jc,%
Damour:2002ws,Blas:2007zz,Carroll:2003wy,Dvali:2008em}, and
Refs.~\cite{Gabadadze:2003ii,Rubakov:2008nh,Sotiriou:2008rp} for
reviews). One of them employs spontaneous breaking of Lorentz symmetry
by space-time dependent condensates of scalar fields
\cite{Arkani-Hamed:2003uy,Dubovsky:2004sg} coupled to gravity in a
covariant way via a derivative coupling. The resulting theory may have
a non-pathological perturbative behavior about the broken vacuum
\cite{Rubakov:2004eb,Dubovsky:2004sg} and exhibit modifications of
gravitational interactions at large scales (see
Sect.~\ref{sc:intro-model} for more details). In particular, graviton
may acquire a non-zero mass, which is the reason to call these models
massive gravity models.

Existing experimental data constrain the mass of the graviton and
other parameters of massive gravity models. Absence of Lorentz
invariance makes the constraints weaker than one would expect in a
Lorentz-invariant theory: the Newton's potential remains
unmodified in the linear approximation despite non-zero graviton mass
\cite{Dubovsky:2005dw}, so that Solar system constraints are satisfied
for rather large masses. The constraint on the mass of the graviton
comes from the emission of gravitational waves by binary pulsars which
is responsible for their spin-down \cite{Taylor:1982zz}. Consistency
of observations with GR requires the graviton mass to be smaller than
the inverse period of orbital motion of the binary system, that is
\cite{Dubovsky:2004ud}
\[
m \lesssim 10^{-19}~{\rm eV}.
\]
Standard cosmological solutions may be reproduced in massive gravity
\cite{Dubovsky:2005dw}. Further constraints on the parameters of the
model are imposed by the growth of perturbations and structure
formation in the post-inflationary epoch \cite{Bebronne:2007qh}.

In General Relativity (GR), a crucial role is played by the
spherically symmetric vacuum solution to the Einstein equations ---
the Schwarzschild solution. This role is twofold: First, this solution
describes the metric outside of spherical non-rotating bodies and
gives rise, in the weak field limit, to the Newtonian gravity. It
provides therefore a useful approximation in many astrophysical
situations.

Second, the Schwarzschild solution describes the result of a
gravitational collapse, the black hole. Although the existence of
black holes has not yet been directly confirmed, there exists an
indirect evidence that some of binary stellar systems contain black
holes as one of the companion
\cite{1986ApJ...308..110M,McClintock:2003gx}, and that many galaxies,
including the Milky Way, harbor super-massive black holes in their
centers \cite{Genzel:1997im,Gillessen:2008qv}. It is conceivable that
black holes will be directly observed in the near future, and that
their properties, including the metric configuration near the horizon,
will be quantitatively tested \cite{Will:2005va}, thus providing a
probe of GR in a fully non-linear regime.

The Schwarzschild metric, together with properly arranged scalar
fields is a solution to Einstein equations in massive gravity as well
\cite{Dubovsky:2007zi}. However, the properties of black holes are, in
general, expected to be different.  In particular, rotating black
holes are certainly modified, and, more generally, black holes are
expected to have hair \cite{Dubovsky:2007zi}. The possible existence
of black hole hair in massive gravity models suggests that there might
exist spherically symmetric solutions other than the Schwarzschild
one. In this paper we construct explicit examples of such solutions.

We found a new class of vacuum spherically-symmetric solutions in
massive gravity which depend, in addition to the mass $M$
(equivalently, the Schwarzschild radius), on one more parameter which
we call the ``scalar charge'' $S$. At zero value of this parameter the
standard Schwarzschild solution is recovered, while at non-zero values
of $S$ the Schwarzschild metric gets modified. The modified solution
is non-linear at all distances; it cannot be obtained in the linear
approximation. Similar phenomenon has been previously found in the
context of bi-gravity models \cite{Berezhiani:2008nr}. The new
solutions may have event horizons and are, therefore, candidates for
modified black holes. We found both analytical and numerical examples
of such modified black holes.

The analytical solutions found in a particular massive gravity model
show a variety of different behaviors. Depending on the parameters of
the model, the deviation of the metric from the Minkowski one may
decay at infinity as $1/r$ or slower. The solutions thus may have
finite or infinite ADM \cite{Arnowitt:1961zz} mass, respectively. In
the case of a finite mass, this mass may be positive or negative
depending on the sign of $M$. In either case the singularity at the
origin may be hidden by the horizon. The solutions with negative mass
exhibit an anti-gravitating behavior at large distances from the
center. 

The modified gravitational potential may decay slower than $1/r$ in a
certain distance range. Such a behavior would mimic the effect of the
dark matter. Interestingly, the modification depends not only on the 
parameters of the model, but also on the scalar charge and the mass of
the solution. Thus, the apparent amount of the ``dark matter'' may be
different for objects of the same mass, in contrast with other models
of modified gravity.

This paper is organized as follows. In Sect.~\ref{sc:intro-model} we
briefly review the massive gravity model and summarize previous
results about the Newtonian potential in this model. In
Sect.~\ref{sc:ansatz} we present the spherically symmetric ansatz and
reduce the Einstein equations to four ordinary differential
equations. In Sect.~\ref{sc:XW} we choose a particular model and find
analytical solutions to the Einstein equations in this model. In
Sect.~\ref{sc:Z} we demonstrate, by numerical computation, that
modified black hole solutions exist in more realistic massive gravity
models which are attractors of the cosmological evolution. Finally,
Sect.~\ref{sc:concl} contains the summary and discussion of our
results.

\section{The massive gravity model} \label{sc:intro-model}

In this paper we consider the massive gravity model described by the
following action \cite{Dubovsky:2004sg},
\begin{eqnarray} 
\label{eq:action}
\mathcal{S} = \int \dif x^{4} \sqrt{- g} 
\left[ - \Mpl \mathcal{R} + \mathcal{L}_{\textrm{m}} + 
\Lam \mathcal{F} \right] .
\end{eqnarray}
The first two terms are the curvature and the Lagrangian of the
minimally coupled ordinary matter; they comprise the standard GR
action. The third term describes four scalar fields $\phi^0$, $\phi^i$
whose space-time dependent vacuum expectation values break
spontaneously the Lorentz symmetry. These fields are minimally coupled
to gravity through a derivative coupling; they will be referred to as
the Goldstone fields. We consider the functions ${\cal F}$ which
depend on two particular combinations of the derivatives of the
Goldstone fields, $\mathcal{F} = \mathcal{F} \left( X, W^{ij}
\right)$, where
\begin{eqnarray*}
X &=& \dfrac{\partial^{\mu} \phi^0 \partial_{\mu} \phi^0}{\Lambda^4} ,\\
W^{ij} &=& \dfrac{\partial^{\mu} \phi^i
\partial_{\mu}\phi^j}{\Lambda^4} 
- \dfrac{\partial^{\mu} \phi^i \partial_{\mu}\phi^0\,
\partial^{\nu} \phi^j \partial_{\nu}\phi^0}{\Lambda^8 X} .
\end{eqnarray*}
The constant $\Lambda$ has the dimension of mass. The model is
understood as the low-energy effective theory valid below the scale
$\Lambda$.

The vacuum configuration has the form  
\begin{eqnarray} \label{eq:vacuum}
g_{\mu\nu} = \eta_{\mu\nu} , & \phi^{0} = a \Lambda^2 t , & \phi^{i} =
b \Lambda^2 x^{i} ,
\end{eqnarray}
where $a$ and $b$ are two constants determined by the requirement that
the energy-momentum tensor associated with the four scalar fields
vanishes in the Minkowski background. The configuration
(\ref{eq:vacuum}) is, therefore, a solution to the Einstein equations.
The constants $a$ and $b$ may be set to one by the redefinition of
fields, which we assume to be the case in what follows. 

For functions $\mathcal{F}$ which are invariant under rotations of the
Goldstone fields $\phi^i$ in the internal space (i.e., those depending
on $W^{ij}$ through three combinations $w_n = {\rm Tr} W^n$,
$n=1,2,3$), the background (\ref{eq:vacuum}) preserves the rotational
symmetry. The Lorentz symmetry is, in general, broken.

The action of the Goldstone fields with functions $\mathcal{F}$
depending only on $X$ and $W^{ij}$ is invariant under the following
symmetry,
\begin{eqnarray*}
\phi^i \rightarrow \phi^i + \chi^i \left( \phi^0 \right) ,
\end{eqnarray*}
where $\chi^i$ are arbitrary functions of $\phi^{0}$. Because of this
symmetry, the behavior of perturbations about the vacuum
(\ref{eq:vacuum}) is non-pathological, i.e. there are neither ghost
nor rapid instabilities \cite{Dubovsky:2004sg}. The spectrum consist
of two tensor modes (graviton polarizations) only, which are, in
general, massive. The graviton mass scale is $m\sim \Lambda^2/M_{\rm
  Pl}$. 

During the cosmological evolution, the Universe described by the
action (\ref{eq:action}) is driven to an ''attractor`` point 
 \cite{Dubovsky:2005dw}
\begin{eqnarray*}
\mathcal{F} \left( X, W^{ij} \right) \rightarrow 
\mathcal{F} \left( Z^{ij} \right)
&\textrm{where}& Z^{ij} = X^{\gamma} W^{ij},
\end{eqnarray*}
$\gamma$ being a constant. At the attractor point the theory
possesses an additional symmetry,
\begin{eqnarray}
\label{eq:symmetry2}
\phi^0 \rightarrow \lambda \phi^0, && \phi^i
\rightarrow \lambda^{-\gamma} \phi^i .
\end{eqnarray}
Models described by the action \refp{eq:action} with the function
$\mathcal{F} = \mathcal{F} \left( Z^{ij} \right)$ have been studied
more intensively
\cite{Dubovsky:2005dw,Bebronne:2007qh,Bebronne:2008tr}. In particular,
it has been shown that for $- 1 < \gamma < 0$ and for $\gamma = 1$ the
cosmological perturbations in these models behave identically to those
in GR \cite{Bebronne:2007qh}. For other values of $\gamma$ the
behavior of the perturbations may or may not reproduce that of GR
depending on the initial conditions.

Another reason to study models characterized by the function
$\mathcal{F} = \mathcal{F} \left( Z^{ij} \right)$ comes from the
analysis of Newtonian approximation. The gravitational potential of a
static source in the model (\ref{eq:action}), in the linear approximation, has the form
\cite{Dubovsky:2005dw}
\begin{eqnarray*}
\Phi = M G_N \left( - \dfrac{1}{r} + \mu^2 r \right),
\end{eqnarray*}
where $G_N = \left( 8 \pi \Mpl \right)^{-1}$ is the Newton's constant
and $\mu$ is a constant of order of the graviton mass whose value
depends on the particular form of the function ${\cal F}$. This
constant vanishes at the point where the symmetry (\ref{eq:symmetry2})
holds, that is, where $\mathcal{F} = \mathcal{F} \left( Z^{ij}
\right)$. Thus, at the point of the attractor the Newtonian potential
remains unmodified.

It has been shown that the standard Schwarzschild metric is a solution
to the Einstein equations in massive gravity models possessing the
symmetry (\ref{eq:symmetry2}) \cite{Dubovsky:2007zi}, with the scalar
fields given by
\begin{eqnarray*}
\nonumber
\phi^0 &=&\Lambda^2 \left( t + 2\sqrt{rr_s} + r_s \ln
    \dfrac{\sqrt{r}-\sqrt{r_s}}{\sqrt{r}+\sqrt{r_s}}\right),\\
\phi^i &=& \Lambda^2x^i,
\end{eqnarray*}
where $r_s$ is the Schwarzschild radius of the black hole. On the
other hand, the metric of a rotating black hole is necessarily
modified \cite{Dubovsky:2007zi} from its standard GR (Kerr) form.

\section{Static spherically symmetric ansatz and equations} 
\label{sc:ansatz}

A static spherically symmetric configuration in the massive gravity
model (\ref{eq:action}) can be
written in the following form,
\begin{eqnarray*}
\dif s^2 &=& \alpha ( r ) \dif t^2 + 2 \delta ( r ) 
\dif t \dif r - \beta ( r ) \dif r^2 - \kappa ( r ) \dif \Omega^2 , 
\nonumber \\
\phi^0 &=& \Lambda^2 \left[ t + h \left( r \right) \right] , \nonumber \\
\phi^i &=& \phi \left( r \right) \dfrac{\Lambda^2 x^i}{r} .
\end{eqnarray*}
This field configuration is invariant under two residual coordinate
transformations. The first one is an arbitrary change of the radial
coordinate $r \rightarrow r^\prime = r^\prime \left( r \right)$, which
allows to set either $\kappa = r^2$ or $\phi = r$. The second one
consist in redefining the time variable $t \rightarrow t^\prime = t +
\tau \left( r \right)$. This last transformation allows to cancel
either $\delta \left( r \right)$ or $h \left( r \right)$. We 
choose the conditions $\kappa = r^2$ and $\delta = 0$. Thus, we get
the following ansatz,
\begin{eqnarray} \label{eq:ansatz}
\dif s^2 &=& \alpha(r) \dif t^2 - \beta(r) \dif r^2 
- r^2 \left( \dif \theta^2 + \sin^2 \theta 
\dif \varphi^2 \right) , \nonumber \\
\phi^0 &=& \Lambda^2 \left[ t + h \left( r \right) \right] , \nonumber \\
\phi^i &=& \phi \left( r \right) \dfrac{\Lambda^2 x^i}{r} .
\end{eqnarray}
As compared to GR, this configuration contains two additional radial
functions $h(r)$ and $\phi(r)$. 

As has been pointed out in Sect.~\ref{sc:intro-model}, the rotational
invariance of the vacuum (and likewise, of the ansatz
(\ref{eq:ansatz})) requires that the function ${\cal F}$ depends on
$W^{ij}$ through three combinations $w_n = {\rm Tr}(W^n)$. These
combinations are expressed in terms of the radial functions 
and their derivatives as follows, 
\begin{eqnarray*} \label{eq:w_n}
w_{1} && = - \left( f_{1} + 2 f_{2} \right) , \nonumber \\
w_{2} && = f_{1}^2 + 2 f_{2}^2 , \nonumber \\
w_{3} && = - \left( f_{1}^3 + 2 f_{2}^3 \right) ,
\end{eqnarray*}
where the two functions $f_{1}$ and $f_{2}$ are 
\begin{eqnarray*}
f_{1} = \dfrac{\phi^{\prime 2}}{\alpha\beta X} ,
&& f_{2} = \dfrac{\phi^{2}}{r^2} ,
\end{eqnarray*}
and 
\begin{eqnarray*}
X &= & \dfrac{\beta - \alpha h'^2}{\alpha\beta}. 
\end{eqnarray*}	
In these expressions and in what follows, the prime denotes the
derivative with respect to the radial coordinate $r$.

After fixing the ansatz (\ref{eq:ansatz}), the Einstein equations
reduce to the following four equations:
\begin{eqnarray*}
\mathcal{G}_{0}^{0} = \dfrac{1}{\Mpl} \mathcal{T}_{0}^{0} , 
&& \mathcal{G}_{r}^{r} = \dfrac{1}{\Mpl} \mathcal{T}_{r}^{r} , \\
\mathcal{G}_{\theta}^{\theta} = 
\dfrac{1}{\Mpl} \mathcal{T}_{\theta}^{\theta}, &&
0 = \mathcal{T}_{0}^{r} ,
\end{eqnarray*}
where $\mathcal{G_{\mu\nu}}$ is the Einstein tensor and
$\mathcal{T}_{\mu}^{\nu}$ is the energy-momentum tensor of the four
Goldstone fields. The other six equations are identically
satisfied. The explicit expressions for the components of
$\mathcal{G}_\mu^\nu$ and $\mathcal{T}_\mu^\nu$ are given in the
Appendix.

Consider first the equation
$\mathcal{T}_{0}^{r} = 0$. Assuming $h' \neq 0$, this equation gives 
\begin{eqnarray}
\label{eq:T_0r=0}
&&0 = X \mathcal{F}_X 
+ f_{1} \left( \mathcal{F}_1 - 2 f_1 \mathcal{F}_2 
+ 3 f_{1}^2 \mathcal{F}_3 \right),
\end{eqnarray}
where $\mathcal{F}_X \equiv \partial \mathcal{F}/ \partial X$ and
$\mathcal{F}_i \equiv \partial \mathcal{F}/ \partial w_i$.
Furthermore, the time and radial components of the energy-momentum
tensor differ by the quantity proportional to
eq.~(\ref{eq:T_0r=0}). Therefore, when this equation holds one has
$\mathcal{T}_{0}^{0} = \mathcal{T}_{r}^{r}$. This implies that
$\mathcal{G}_{0}^{0} = \mathcal{G}_{r}^{r}$ or, equivalently,
\begin{eqnarray*}
&& \alpha 
\left( r \right) \beta \left( r \right) = 1 ,
\end{eqnarray*}
in full analogy with the Schwarzschild solution in GR. 
Hence, the Einstein equations reduce to the following four equations,
\begin{eqnarray}
1 &=& \alpha \beta , \label{eq:equat1} \\
0 &=& \dfrac{\alpha^{\prime}}{r} + \dfrac{\alpha - 1}{r^2} 
- \dfrac{m^2}{2} \left( \mathcal{F} 
- 2 X \mathcal{F}_X \right) ,  
\label{eq:equat2} \\
0 &=& \dfrac{\alpha^{\prime}}{r} + \dfrac{\alpha^{\prime\prime}}{2} 
- \dfrac{m^2}{2} \left[ \mathcal{F} 
+ X \mathcal{F}_X \right. \nonumber \\
& & \left. - w_{1} \mathcal{F}_{1}- 2 w_{2} \mathcal{F}_{2} -
3 w_{3} \mathcal{F}_{3}  \right]
, \label{eq:equat3} \\
0 &=& X \mathcal{F}_X 
+ f_{1} \left( \mathcal{F}_1 - 2 f_1 \mathcal{F}_2 
+ 3 f_{1}^2 \mathcal{F}_3 \right),
\label{eq:equat4}
\end{eqnarray}
where $m^2 = \Lam / \Mpl$. For a generic function $\mathcal{F}$, this
system of equations is well defined. Indeed, since the function $h(r)$
enters the equations (\ref{eq:equat1}--\ref{eq:equat4}) only through
the variable $X$, one may consider $X$ as an independent variable
instead of $h(r)$. Then the fourth equation allows to find $\phi$ in
terms of $X$, while the first equation gives $\beta$ in terms of
$\alpha$. The second equation then gives $X$ in terms of $\alpha$ and
the third equation allows to determine $\alpha$ as a function of $r$.

\section{Analytical example} 
\label{sc:XW}

Finding analytical solutions of the non-linear system of equations
like (\ref{eq:equat1}--\ref{eq:equat4}) is impossible for a generic
function ${\cal F}$. So, in order to get some insight into the
behavior of the solutions, let us choose the function $\mathcal{F}$ in
such a way that the resulting equations are solvable analytically.

Consider the function $\mathcal{F}$ of the following form,
\begin{eqnarray} \label{eq:fct-XW}
\mathcal{F} &=& c_0 \left( \dfrac{1}{X} +w_1 \right) 
\\ \nonumber &&
+ c_1\left( w_{1}^3 - 3 w_{1} w_{2} -6w_1
+ 2 w_{3}  -12\right),
\end{eqnarray}
where $c_0$ is an arbitrary dimensionless constant and $c_1=\pm 1$
(the numerical value of $c_1$ can be absorbed into the constant
$\Lambda$). The coefficients inside the parentheses are chosen in such
a way that the vacuum (\ref{eq:vacuum}) is the solution to the
Einstein equations at $a=b=1$. Our example contains, therefore, a
single continuous free parameter $c_0$. Two additional constraints
should be imposed on $c_0$. The first one comes from
the requirement that the graviton is non-tachyonic. This translates
into the inequality
\begin{equation}
c_0-6c_1\geq 0. 
\label{eq:positive-mass}
\end{equation}
The second condition is necessary to ensure that scalar modes with
pathological behavior do not reappear upon addition of
higher-derivative terms (recall that the model (\ref{eq:action}) is
understood as the low-energy effective theory, so such terms are
generically present). This imposes certain constraint on the function
${\cal F}$ and its derivatives (cf. eq.~(70) of
Ref.\cite{Dubovsky:2004sg}), which in the case at hand implies
\begin{equation}
c_0>0.
\label{eq:no-tachyons}
\end{equation}

We should stress that the choice of the functional form of
$\mathcal{F}$ is by no means unique. The particular form
(\ref{eq:fct-XW}) has been chosen in order to simplify the solution of
the field equations, as will become clear below.

\subsection{The static spherically symmetric solution}

Let us start with eq.~\refp{eq:equat4} which takes the form 
\begin{eqnarray*}
0 = \dfrac{1}{X} \left[ - c_{0} + \phi'^2 
\left( 6 c_1 \dfrac{\phi^{4}}{r^4} + c_0 - 6c_1  \right) \right] .
\end{eqnarray*}
Because of our choice of the function $\mathcal{F}$, this equation
contains the variable $X$ as an overall factor only. Hence, it reduces
to a closed differential equation for $\phi$. 
The solution to this equation is 
\[
\phi = b r,
\]
where the
constant $b$ satisfies the equation
\begin{eqnarray} \label{eq:a}
0 = (b^2-1) ( 6b^4 + 6 b^2 + c_0/c_1).
\end{eqnarray}
We are interested in real positive values of $b$ (the
case $b<0$ can be reduced to $b>0$ by the inversion of
coordinates). If $c_0/c_1>0$ there is only one such solution, 
\[
b=1.
\]
If $c_0/c_1<0$ there exists another one, 
\[
b=\dfrac{1}{\sqrt{2}}\left( -1+\sqrt{1-2c_0/3c_1}\right)^{1/2}.
\]
Thus, at negative $c_0/c_1$ we have two different branches of
solutions. 

The remaining equations \refp{eq:equat2} and \refp{eq:equat3} can be
written as follows,
\begin{eqnarray}\nonumber
0 &=& \dfrac{\alpha^{\prime}}{r} + \dfrac{\alpha - 1}{r^2} 
- 3 \Lambda_{c} + m^2 c_0
\left( 1 - \dfrac{1}{X} \right) , \\ \label{eq:alpha(r)}
0 &=& \alpha^{\prime\prime} + \lambda \dfrac{\alpha - 1}{r^2} 
+ \left( \dfrac{\alpha^{\prime}}{r} 
- 3 \Lambda_{c} \right) \left( 2 + \lambda \right) ,
\end{eqnarray}
where 
\begin{eqnarray*}
\label{eq:lambda}
\lambda &=& - 12 b^6 \dfrac{c_1}{c_{0}},
\\ 
\label{eq:Lambda}
\Lambda_{c} &=& 2m^2c_1 (b^{6}-1).
\end{eqnarray*}
Recall that, according to our normalization, the constant
$c_1$ only takes two values, $c_1=\pm 1$.

Eq.~(\ref{eq:alpha(r)}) is a linear inhomogeneous equation for
$\alpha$. Its general solution can be found analytically. Making use
of this solution and integrating the remaining equations one finally
obtains:
\begin{eqnarray} \label{eq:sol-XW}
\alpha \left( r \right) &=& 1 - \dfrac{r_{s}}{r} -
\dfrac{S}{r^{\lambda}} + \Lambda_{c} r^2 , \\
\beta \left( r \right) &=&1/ \alpha (r), \nonumber \\
h \left( r \right) &=& \pm \int \dfrac{\dif r}{\alpha} 
\left[ 1 - \alpha \left( \dfrac{S}{c_{0} m^2} 
\dfrac{\lambda-1}{r^{\lambda+2}} + 1 \right)^{-1} 
\right]^{1/2}, \nonumber \\
\phi \left( r \right) &=& b r. \nonumber
\end{eqnarray}
Here $r_{s}$ and $S$ are two integration constants: $r_{s}$ is the
Schwarzschild radius, while $S$ is a scalar charge whose presence
reflects the modification of the gravitational interaction as compared
to GR. At $S=0$ the solution (\ref{eq:sol-XW}) reduces to the
conventional Schwarzschild solution describing  a black hole of the
mass $M=r_s/(2G_N)$. 

The behavior of the metric at $r\to\infty$ depends on the constants
$c_0$ and $c_1=\pm 1$. If $c_1=1$, we must also take $c_0\geq 6$, and
the only solution of eq.~(\ref{eq:a}) in this case is $b=1$. Hence, we
have $\Lambda_c=0$, $\lambda<0$, and the metric is growing at infinity
as $S r^{|\lambda|}$. Such solutions do not describe asymptotically
flat space.

If $c_1=-1$, we must take $c_0>0$ in order to satisfy
eqs.~(\ref{eq:positive-mass}) and (\ref{eq:no-tachyons}). In this case
$\lambda >0$. Two branches of solutions exist, one with zero, and one
with non-zero cosmological constant, which can be positive or negative
depending on the numerical value of $c_0$. In what follows we consider
the branch with $\Lambda_c=0$ ($b=1$).

The behavior of the solutions is determined by the two integration
constants $M$ and $S$, and the value of the parameter $\lambda$. If
$\lambda<1$, the third term on the r.h.s. of eq.~(\ref{eq:sol-XW})
dominates at large distances. The ADM mass of these solutions is
infinite. We do not discuss them further here. 

If $\lambda>1$, the standard Schwarzschild term dominates at
infinity. The ADM mass of such solutions is equal to $M$. The
solutions with positive (negative) $M$ have attractive (repulsive)
behavior at infinity.

\begin{figure*}
\includegraphics[angle=0,width=500pt,height=150pt]{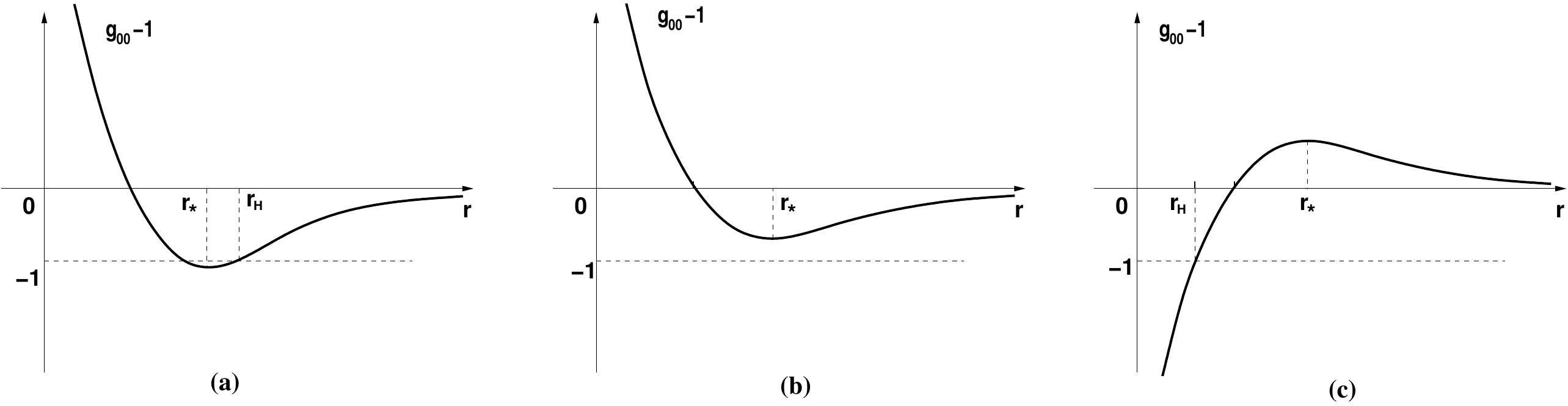}
\caption{The deviation of $g_{00}$ from one (proportional to the
Newtonian potential $\Phi = (g_{00}-1)/2$) for tree different choices
of the integration constants. Figs.~(a) and (b) correspond to $M >
0$ and $S < 0$ with the numerical values satisfying (a) and not
satisfying (b) eq.~(\ref{eq:horizon}). Fig.~(c) represents solution
with $M < 0$ and $S > 0$.}
\label{fig:1}
\end{figure*}

At the origin $r=0$ both terms proportional to $M$ and $S$ are
singular, so the metric always possesses a singularity unless
$M=S=0$. This singularity may or may not be hidden by the horizon
depending on the signs and values of $M$ and $S$. The solutions
possessing the horizon are candidates for modified black holes. 

The horizon is always present if both $M$ and $S$ are positive. Such
black holes have attractive gravitational potential at all distances,
which is stronger than for a conventional black hole of the mass
$M$. The horizon size of the modified black hole is larger than
$r_s=2G_NM$.

If $M>0$ and $S<0$, the presence of the horizon depends on the
relative values of $S$ and $M$. It exists for sufficiently small
$S$. Defining, on dimensional grounds, the mass parameter $s$ associated
with $S$ by the relation $|S|=s^{-\lambda}$, the existence of horizon
requires that
\begin{equation}
s M \geq \dfrac{\lambda}{2G_N}\left(\dfrac{1}{\lambda-1}
\right)^{\dfrac{\lambda-1}{\lambda}}. 
\label{eq:horizon}
\end{equation}
The Newtonian potentials for solutions satisfying and not satisfying
the condition (\ref{eq:horizon}) are shown in Fig.~\ref{fig:1}--(a)
and (b), respectively. When the horizon exists, the gravitational
field is attractive all the way to the horizon. The attraction is
weaker than in the case of the usual Schwarzschild black hole of mass
$M$, and the horizon size is smaller. The behavior of the
gravitational force with distance mimics that of the smaller-mass
black hole plus a continuous distribution of ``dark matter'', with the
total mass enclosed within the radius $r$ approaching $M$ as
$r\to\infty$.

Finally, at $M<0$ and $S>0$ the modified black hole anti-gravitates at
large distances and gravitates close to the horizon. The attraction
changes to repulsion at
\[
r=r_* \equiv  \left|\dfrac{\lambda S}{r_s}\right|^{\dfrac{1}{\lambda-1}}.
\]
The corresponding Newtonian potential is shown in
Fig.~\ref{fig:1}--(c).

A remark is in order at this point. In the conventional general
relativity the constant $r_s$ or equivalently, the black hole mass
$M$, is also a free parameter which can, in principle, be positive or
negative. In GR, however, only positive values make sense for the
following reasons.  First, negative-mass Schwarzschild solutions
possess naked singularity at the origin, which is physically
unacceptable. Second, the conventional matter satisfies the null
energy condition which ensures that any compact spherically-symmetric
matter distribution has a positive mass \cite{Hawking:1973uf}. None of
these arguments go through in the case of massive
gravity. Fig.~\ref{fig:1}--(c) gives an example of solution with
repulsive behavior at large distances and without naked singularity:
as for a conventional black hole, the singularity of this solution is
hidden behind the horizon.  The positivity of energy is also not
expected in massive gravity. This is related to the fact that the
background (\ref{eq:vacuum}) breaks time translations, and only the
combination of the time translations with the shifts of $\phi^0$ by a
constant remains unbroken. In this respect the massive gravity model
is exactly analogous to the ghost condensate model
\cite{Arkani-Hamed:2003uy}, where the negative-energy states have been
constructed explicitely \cite{Arkani-Hamed:2005gu}.

\subsection{Correspondence with linear analysis}

The solutions found in the previous section have the asymptotic
behavior different from that obtained in the linear perturbation
theory in Ref.~\cite{Dubovsky:2004ud}. In order to compare our results
with those of Ref.~\cite{Dubovsky:2004ud}, let us discuss them in the
gauge where $h(r)=0$, $\delta(r)\neq 0$
(cf. Sect.~\ref{sc:ansatz}). In this gauge the perturbation theory of
Ref.~\cite{Dubovsky:2004ud} corresponds to assuming that the
variations of all the metric components are of the same order. In
other words, they are formally assigned a small parameter $\epsilon$
to the first power. The solutions described in
Ref.~\cite{Dubovsky:2004ud} satisfy the Einstein equations expanded to
the linear order in $\epsilon$.

The solution (\ref{eq:sol-XW}) is not of this type. Transforming it
into the gauge $h(r)=0$ one finds that $g_{tr}=\delta(r)$ does not
decay as fast as the perturbations of other components, for instance,
as $\alpha-1$. In fact, in the equations expanded in powers of
perturbations the terms of order $\delta^2$ balance those linear in
$\alpha-1$. In other words, in the formal expansion of the solution of
Sect.~\ref{sc:XW} in powers of the small parameter $\epsilon$ the
perturbation $\delta$ should be assigned the order $\sqrt{\epsilon}$
rather than $\epsilon$. Hence, the solution (\ref{eq:sol-XW}) is
non-linear even at large distances from the center.  A similar
phenomenon has been observed in the context of bi-gravity models in
Ref.~\cite{Berezhiani:2008nr}.

Another difference between the solution (\ref{eq:sol-XW}) and the
solution to the linearized equations of Ref.~\cite{Dubovsky:2004ud} is
that the former is static, while in the latter only metric components
are static (in the gauge $g_{0i}=0$). The scalar fields have time
dependence which may be viewed as an accretion of a fluid with
zero energy-momentum tensor.

\subsection{Gravitational field of a star}
\label{sec:grav-field-star}

In general relativity, one may relate the mass of a star to an
integration constant of the vacuum solution in the exterior space by
matching the interior and exterior solutions at the star surface (see,
e.g., Ref.~\cite{Weinberg:1972}). In massive gravity, one may try to
use the same approach to determine the scalar charge of an ordinary
star. The analytical solution in the interior region is required for
the matching procedure.

The star is described, to a good approximation, by a diagonal
energy-momentum tensor $t_{\mu}^{\nu} = \left( \rho, -p, -p, -p
\right)$, where $\rho$ and $p$ are the energy density and pressure
inside the star, respectively. This energy-momentum tensor is assumed
to be responsible for the external gravitational field described by
eq.~\refp{eq:sol-XW}. Since there is no direct coupling between the
ordinary matter and the Goldstone fields, $t_{\mu}^{\nu}$ must be
conserved separately, $\nabla^\mu t_{\mu}^{\nu}=0$. For simplicity, we
take the energy density to be constant at $r<R$, where $r=R$ is the
surface of the star, and zero outside. The pressure $p$ cannot be
chosen independently; it is determined by the conservation of
$t_{\mu}^{\nu}$.

Because of the spherical symmetry, the ansatz \refp{eq:ansatz}
holds. The Einstein equations in the interior of the star are obtained
from eqs.~(\ref{eq:equat1}--\ref{eq:equat3}) by adding the
contributions of the energy-momentum of the star, while
eq.~(\ref{eq:equat4}) remains unchanged. The resulting set of
equations can be solved analytically. The solution reads
\begin{eqnarray*}
\alpha \left( r \right) &=& 1 - \dfrac{s_{1}}{r} -
\dfrac{s_{2}}{r^{\lambda}} + \Lambda_{c} r^2 \\
& & + \dfrac{\rho}{\Mpl} \left( \dfrac{r^2}{6} - \dfrac{R^2}{2}
\right) 
+ \mathcal{O} \left( \rho^2 \right), \\
\beta \left( r \right) &=& \left[ 1 - \dfrac{s_{1}}{r} -
\dfrac{s_{2}}{r^{\lambda}} + \Lambda_{c} r^2 
- \dfrac{r^2 \rho}{3 \Mpl} \right]^{-1}, \nonumber \\
h \left( r \right) &=& \pm \int \dfrac{\dif r}{\alpha} 
\left[ 1 - \alpha \left( \dfrac{s_{2}}{c_{0} m^2} 
\dfrac{\lambda-1}{r^{\lambda+2}} + 1 \right)^{-1} 
\right]^{1/2}, \nonumber \\
\phi \left( r \right) &=& b r. \nonumber
\end{eqnarray*}
For simplicity, we have expanded the first equation in powers of
$\rho$, while the other relations are exact. Since the geometry inside
the star is regular, the integration constants $s_{1}$ and $s_{2}$
must be set to zero.

The interior solution has to be matched with the solution
(\ref{eq:sol-XW}) at $r=R$. It is convenient to match the variable $X$
which equals 1 in the interior region. In the gauge $h(r)=0$ this
variable is nothing but the $g^{00}$ component of the metric. Hence,
it must be continuous.  Making use of eqs.~(\ref{eq:sol-XW}) one can
see that the continuity of $X$ at $r=R$ requires that
$S=0$. Therefore, the scalar charge of an ordinary star is zero.

It remains an open question how objects (e.g., black holes) with
$S\neq 0$ can be created. The argument given above does not apply to
time-dependent configurations, so it is possible that a non-zero
scalar charge may be acquired during the gravitational collapse.

\section{\texorpdfstring{$\mathcal{F}(Z^{ij})$}{} models} 
\label{sc:Z}

As mentioned earlier, models characterized by the function ${\cal F}$
of a single variable $Z^{ij} = X^\gamma W^{ij}$ are of a particular
interest. We discuss in this section the exact static spherically
symmetric solutions in these models. Our goal is to demonstrate that
the solutions found earlier are not specific to the particular form of
the action (\ref{eq:fct-XW}) and exist also in the models obeying the
symmetry (\ref{eq:symmetry2}).

In Sect.~\ref{sc:XW} the analytical solutions of
eqs.~(\ref{eq:equat1}--\ref{eq:equat4}) were obtained by choosing
the function ${\cal F}$ in such a way that the dependence on $X$
factors out in eq.~(\ref{eq:equat4}). Since now $\mathcal{F}$ has only
one argument, the derivatives of $\mathcal{F}$ with respect to $X$ and
$W^{ij}$ are no more independent. For this reason we did not succeed
in constructing non-trivial examples where the Einstein equations are
solvable analytically. Hence, to demonstrate the existence of
solutions we have to use numerical methods.

Consider the following function $\mathcal{F}$,
\begin{eqnarray} \label{eq:functZ}
\mathcal{F} &=& c_{0} \left( z_{1} + 2 + \dfrac{z_{1}^3 
- 6 z_{1} z_{2} + 8 z_{3}}{3} \right) \\
& & + 2 c_{1} \left( z_{1}^2 - 2 z_{2} - 1 
- 2 \dfrac{z_{1}^3 - 6 z_{1} z_{2} + 8 z_{3}}{3} \right), \nonumber
\end{eqnarray}
were $z_{n} = \textrm{Tr} \left( Z^n \right)$ are tree independent
scalars made out of $Z^{ij}$, $z_{n} = \left( X^\gamma \right)^{n}
w_{n}$. The coefficients in front of individual terms
have been adjusted so that the flat metric and the scalar fields given
by eq.~(\ref{eq:vacuum}) solve the field equations at $a=b=1$.  For
the vacuum configuration (\ref{eq:vacuum}) one has $Z^{ij} = -
\delta^{ij}$. We are interested in solutions to the field equations
that asymptote to this vacuum state. 

In addition to the adjustments already made, the following inequality
should be imposed on the coefficients $c_0$ and $c_1$ to ensure that
the graviton is non-tachyonic,
\begin{equation}
c_0 - 2 c_1 \geq 0.
\label{eq:positive-mass2}
\end{equation}
This guarantees that the square of the graviton mass is non-negative.
Moreover, this inequality is sufficient for the absence of
pathological scalar modes which may appear upon addition of
higher-derivative terms.  As in the previous example, the overall
scale of the coefficients $c_0$ and $c_1$ can be absorbed in the
parameter $\Lambda$, so without loss of generality we may set $c_1=\pm
1$.

For this class of models, the field equations
(\ref{eq:equat1}--\ref{eq:equat4}) may be viewed as equations for
$\alpha(r)$, $\beta(r)$, $\xi(r) \equiv X^\gamma f_{1} - 1$ and
$\psi(r) \equiv X^\gamma f_{2} - 1$. Then eq.~\refp{eq:equat1} gives
$\beta$ in terms of $\alpha$, while eq.~\refp{eq:equat4} enables to
express $\psi$ in terms of $\xi$. The two remaining equations form
a coupled set of non-linear equations for $\alpha$ and $\xi$; they
have to be solved numerically.

\begin{figure}
\includegraphics[angle=0,width=8cm]{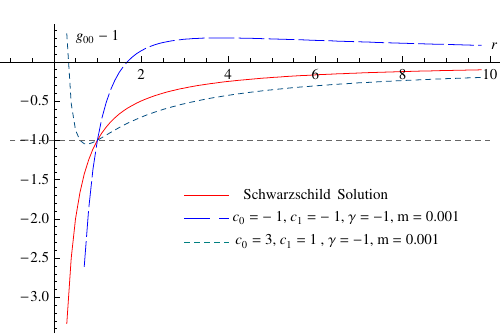}
\caption{The deviation of $g_{00}$ from one for three different cases:
the usual Schwarzschild solution (solid line) and two solutions
corresponding to different values of the parameters of the function
\refp{eq:functZ} (long-dashed and short-dashed lines). The integration
constants of these solutions have been chosen such that the external
horizon is located at $r = 1$. }\label{fg:alpha}
\end{figure}

The numerical solutions are shown in Fig.~\ref{fg:alpha} for different
value of the parameters $c_{0}$ and $c_{1}$. For all these graphs, we
have assumed that the external horizon is located at $r = 1$ and that
$\xi = 100$ at the horizon. The large value of $\xi$ is chosen in
order to make the difference between the modified solution and the
Schwarzschild solution visible on the plot (large values of $\xi$
correspond to large scalar charge $S$ of the previous section).

The plots show the behavior qualitatively similar to that discussed in
Section \ref{sc:XW}. In particular:
\begin{itemize}
\item if $c_0 = c_1 = - 1$, the Newtonian potential $2 \Phi = \alpha -
1$ is attractive at short distances and becomes repulsive at larger
distances;
\item if $c_0 = 3$ and $c_{1} = 1$, the Newtonian potential is attractive
outside of the horizon, and becomes repulsive close to the
singularity.
\end{itemize}
The deviations from the Schwarzschild metric are larger for larger
values of the ``scalar charge'' (parameterized by the value of $\xi$
at the horizon). The Schwarzschild solution is recovered at $\xi\to
0$.

\section{Discussion}
\label{sc:concl}

To summarize, there exist spherically symmetric vacuum solutions in
massive gravity models which depend on two integration constants, the
mass $M$ and an extra parameter $S$ which can be called the ``scalar
charge''. At zero value of the scalar charge the standard
Schwarzschild solution is recovered, while at non-zero $S$ the metric
is modified with respect to the Schwarzschild case.

The solutions having non-zero scalar charge exhibit much reacher
behavior than the Schwarzschild solution in GR. As can be seen
from the explicit example of Sect.~\ref{sc:XW}, both the short and
long distance behavior may be modified depending on the parameters of
the model. 

Unlike in General Relativity, the solutions may have a {\em negative}
ADM mass. Such solutions have repulsive gravitational interaction at
large distances. At short distances the repulsion may change to
attraction and give rise to the horizon, hiding the singularity at the
origin. Such solutions represent anti-gravitating black holes.

In the case of a positive ADM mass, the $S$-dependent contributions
may make the gravitational attraction weaker at short distances
(cf. fig.~\ref{fig:1}-(a)). In this case the gravitational force
decays with distance slower than $1/r^2$, thus mimicking the presence
of dark matter. Interestingly, solutions with the same value of $M$
but different scalar charge $S$ have different behavior, which
corresponds to different amount of the apparent ``dark matter''. This
is in contrast with other models possessing modifications of the
gravitational potential
\cite{Milgrom:1983pn,Dvali:2000hr,Carroll:2003wy}, where the
modification of the gravitational force is determined by the
parameters of the model.

It is currently an open question how objects with non-zero scalar
charge may be created. As has been argued in
Sect.~\ref{sec:grav-field-star}, the absence of direct coupling
between the Goldstone fields and ordinary matter results in zero
scalar charges of static matter distributions. Thus, the gravitational
field of ordinary stars is described by the $S=0$ solutions, i.e., by
the standard Schwarzschild metric. This may be not the case for black
holes, especially the super-massive black holes in the centers of
galaxies, which may be of primordial origin
\cite{Carr:1974nx,Carr:2005bd}. In any case, this question requires
further investigation.

Another open question is the stability of the modified black hole
solutions.  Several kinds of instabilities may be present. Among
perturbations of the solutions there may exist unstable modes with the
characteristic time scale of order of the horizon size; in this case
the interpretation in terms of black holes is not possible. Second
potential source of problems is generic presence of the
higher-derivative terms not included in the action
(\ref{eq:action}). One has to check that the 2-parameter family of
modified black holes survives their inclusion. By analogy with the
ghost condensate case, one may expect that these terms produce at
least a slow Jeans-type instability \cite{Arkani-Hamed:2003uy}, which
is not, however, dangerous for the black hole interpretation. Finally,
the presence of negative mass solutions may lead to instabilities of
the quantum-mechanical nature similar to those found in
Ref.\cite{Krotov:2004if}.

To conclude this list, let us mention also the solutions satisfying $h
= 0$ which were not considered in this paper. In this case the ten
Einstein equations reduce to three equations for $\alpha$, $\beta$ and
$\phi$ which form a (generically) well-defined system. It remains to
be seen whether this system has asymptotically flat solutions. In any
case, the Schwarzschild solution does not belong to this class which
is characterized by $\alpha \beta \neq 1$.

\section*{Acknowledgments}

We are grateful to S. Dubovsky for valuable discussions and comments on the manuscript.
The work of M.B. is supported by the Belgian \emph{Fond pour la
Formation \`a la Recherche dans l'Industrie et dans l'Agriculture
(FRIA)}. The work of P.T. is supported by IISN, Belgian Science Policy
(under contract IAP V/27).

\appendix

\section*{Appendix}

For the ansatz \refp{eq:ansatz} discussed in Section \ref{sc:ansatz},
the non-zero components of the Einstein tensor are given by the
following relations
\begin{eqnarray*}
\mathcal{G}_{0}^{0} &=& \dfrac{1}{r^2} \left[ 1 - \left( \dfrac{r}{\beta} \right)^{\prime} \right], \\
%%%
\mathcal{G}_{r}^{r} &=& \dfrac{1}{r^2} \left( 1 - \dfrac{\alpha + r \alpha^{\prime}}{\alpha \beta} \right), \\
%%%
\mathcal{G}_{\theta}^{\theta} &=& \mathcal{G}_{\varphi}^{\varphi} = - \dfrac{1}{4 r} \left[ \dfrac{\alpha^{\prime} + r \alpha^{\prime\prime}}{\alpha \beta} + \left( \dfrac{2 \alpha + r \alpha^{\prime}}{\alpha \beta} \right)^{\prime} \right].
\end{eqnarray*}
The energy-momentum tensor of the Goldstone fields is expressed
through
\begin{eqnarray*}
\mathcal{T}_{\mu}^{\nu} &=& - \dfrac{1}{2} \Lambda^{4} \delta_{\mu}^{\nu} \mathcal{F} + \mathcal{F}_{X} \partial^\nu \phi^0 \partial_\mu \phi^0 \\
& & + \dfrac{\partial \mathcal{F}}{\partial W^{ij}} \left[ \partial^\nu \phi^i \partial_\mu \phi^j + \dfrac{V^{i} V^{j}}{X^{2}} \partial^\nu \phi^0 \partial_\mu \phi^0 \right. \\
& & \left. - \dfrac{V^{j}}{X} \left( \partial^\nu \phi^i \partial_\mu \phi^0 + \partial^\nu \phi^0 \partial_\mu \phi^i \right) \right],
\end{eqnarray*}
where $\mathcal{F}_{X} \equiv \partial \mathcal{F} / \partial X$ and
\begin{eqnarray*}
V^{i} = \dfrac{\partial^\mu \phi^0 \partial_\mu \phi^i}{\Lambda^4}.
\end{eqnarray*}
For functions $\mathcal{F}$ which are invariant under rotations of the Goldstone fields $\phi^i$ internal space, the derivatives of $\mathcal{F}$ with respect to $W^{ij}$ are given by
\begin{eqnarray*}
\dfrac{\partial \mathcal{F}}{\partial W^{ij}} &=& \mathcal{F}_{1} \delta_{ij} + 2 \mathcal{F}_{2} W^{ij} + 3 \mathcal{F}_{3} W^{ik} W^{kj},
\end{eqnarray*}
where $\mathcal{F}_{i} \equiv \partial \mathcal{F} / \partial w_{i}$. Therefore, the components of the energy-momentum tensor which are not identically zero are given by
\begin{eqnarray*}
\mathcal{T}_{0}^{0} &=& \Lambda^{4} \left[ - \dfrac{1}{2} \mathcal{F} + \dfrac{1}{\alpha} \left( \mathcal{F}_{X} + \dfrac{\partial \mathcal{F}}{\partial W^{ij}} \dfrac{V^{i} V^{j}}{X^{2}} \right) \right] , \\
& & \\
\mathcal{T}_{r}^{r} &=& \Lambda^{4} \left[ - \dfrac{1}{2} \mathcal{F} - \left( \mathcal{F}_{X} + \dfrac{\partial \mathcal{F}}{\partial W^{ij}} \dfrac{V^{i} V^{j}}{X^{2}} \right) \dfrac{h^{\prime2}}{\beta} \right] \\
& & + \dfrac{1}{\beta} \dfrac{\partial \mathcal{F}}{\partial W^{ij}} \left( - \partial_r \phi^i \partial_r \phi^j + \dfrac{2 V^{j}}{X} \partial_r \phi^i \partial_r \phi^0 \right)  , \\
\mathcal{T}_{\theta}^{\theta} &=& \mathcal{T}_{\varphi}^{\varphi} = - \dfrac{1}{2} \Lambda^{4} \mathcal{F} + \dfrac{\partial \mathcal{F}}{\partial W^{ij}} \partial^\theta \phi^i \partial_\theta \phi^j , \\
\mathcal{T}_{0}^{r} &=& - \dfrac{\Lambda^{4} h^{\prime}}{\beta} \left[ \mathcal{F}_{X} + \dfrac{\partial \mathcal{F}}{\partial W^{ij}} \left( \dfrac{V^{i} V^{j}}{X^{2}} + \dfrac{\partial_r \phi^i \partial_r \phi^{j}}{\beta \Lambda^{4} X} \right) \right].
\end{eqnarray*}

\end{document}